\documentclass{aa}
\usepackage{txfonts}
\usepackage{graphicx}
\usepackage{natbib}
\bibpunct{(}{)}{;}{a}{}{,} 

\begin{document}

\title{Quasars in the MAMBO blank field survey\thanks{Based on
observations with the 100-m telescope of the MPIfR
(Max-Planck-Institut f\"ur Radioastronomie) at Effelsberg, with the IRAM
Plateau de Bure Interferometer and the IRAM 30m MRT at Pico Veleta, with
the NRAO VLA, with the Spitzer Space Telescope, which is operated by the
Jet Propulsion Laboratory, California Institute of Technology under NASA
contract 1407, and with the Gemini Observatory, which is operated by the
Association of Universities for Research in Astronomy, Inc., under a
cooperative agreement with the NSF on behalf of the Gemini partnership: the
National Science Foundation (United States), the Particle Physics and
Astronomy Research Council (United Kingdom), the National Research Council
(Canada), CONICYT (Chile), the Australian Research Council (Australia),
CNPq (Brazil) and CONICET (Argentina). IRAM is supported by  INSU/CNRS
(France), MPG (Germany) and IGN (Spain). The National Radio Astronomy
Observatory (NRAO) is a facility of the National Science Foundation
operated under cooperative agreement by Associated Universities, Inc.}}

\author{Hauke Voss      \inst{1} 
\and    Frank Bertoldi  \inst{1,2} 
\and    Chris Carilli   \inst{3} 
\and    Frazer N. Owen  \inst{3} 
\and    Dieter Lutz     \inst{4} 
\and    Mark Holdaway   \inst{3} 
\and    Michael Ledlow  \inst{5}\thanks{Deceased 5 June 2004. We shall miss his 
cheerfulness, unfailing good sense and scientific industry.}
\and    Karl M. Menten  \inst{1}
}

\institute{Max-Planck Institut f\"ur Radioastronomie, Auf dem H\"ugel
69, D-53121 Bonn, Germany 
\and Radioastronomisches Institut der Universit\"at Bonn, Auf dem H\"ugel 71, 
D-53121 Bonn, Germany 
\and National Radio Astronomy Observatory, P.O. Box, Socorro, NM 87801, USA 
\and Max-Planck Institut f\"ur extraterrestrische Physik, Giessenbachstra{\ss}e, 
D-85748 Garching, Germany 
\and Gemini Observatory, Southern Operations Center, AURA, Casilla 603, La Serena Chile
}

\offprints{Hauke Voss,\\ \email{hvoss@mpifr-bonn.mpg.de}}
\date{Received date / Accepted date}

\abstract{Our MAMBO 1.2 mm blank field imaging survey of $\sim 0.75$ sqd has uncovered four
unusually bright sources, with flux densities between 10 and 90 mJy, all located in the
Abell~2125 field. The three brightest are flat spectrum radio sources with bright optical and
X-ray counterparts. Their mm and radio flux densities are variable on timescales of months. Their
X-ray luminosities classify them as quasars. The faintest of the four mm~bright sources appears
to be a bright, radio-quiet starburst at $z\sim3$, similar to the sources seen at lower flux
densities in the MAMBO and SCUBA surveys. It may also host a mildly obscured AGN of quasar-like
X-ray luminosity. The three non-thermal mm sources imply an areal density of flat spectrum radio
sources higher by at least 7 compared with that expected from an extrapolation of the lower
frequency radio number counts.

\keywords{Galaxies: active -- Galaxies: high redshift -- Galaxies: starburst 
-- Galaxies: individual: MMJ1540+6605, MMJ1541+6630, MMJ1541+6622, MMJ1543+6621 
-- quasars: general -- Submillimeter}}

\maketitle

\section{Introduction}

Blank field surveys with the SCUBA and MAMBO bolometer arrays were able to
resolve a good fraction of the extragalactic far-infrared (FIR) background
radiation that was discovered by COBE \citep{Puget96} into point sources,
revealing a population of previously unknown dust-enshrouded, optically faint
starburst galaxies at high redshifts \citep{Smail97, Hughes98, Barger98,
Bertoldi02a, Blain02, Chapman05}. One aim of the MAMBO survey was to cover a large field, in order
to better constrain the abundance and nature of the brightest mm background
sources. By now, about 2700~arcmin$^2$ have been imaged to rms depths ranging from
0.5 to 3~mJy per 11~arcsec beam \citep{Bertoldi00, Dannerbauer02, Dannerbauer04,
Eales03, Greve04}. Most of the MAMBO and SCUBA sources show faint radio (20~cm)
counterparts. Their radio-to-mm spectral energy distributions (SEDs) are similar to
starburst galaxies, which are dominated in the radio by synchrotron emission,
and in the (sub)mm by a steeply rising thermal dust emission spectrum.

The estimated infrared luminosities of the (sub)mm background sources imply
star formation rates of order 1000~M$_\odot$~yr$^{-1}$, which to sustain
would require them to be very massive systems, comparable to the massive
spheroids in the local universe. Recent CO detections in seven SCUBA and MAMBO
galaxies imply luminous and dynamical masses of order $10^{11}$~M$_\odot$
\citep{Neri03, Greve05}.

It has been a main motivation for the wide field MAMBO surveys to constrain the
volume density of the most massive starburst galaxies at high redshifts. We
here report first results of our shallow wide field surveys, focusing on the
detection of the four brightest sources ever found in mm or submm blank field surveys.

\section{Observations}

\begin{figure}
\resizebox{\hsize}{!}{\includegraphics{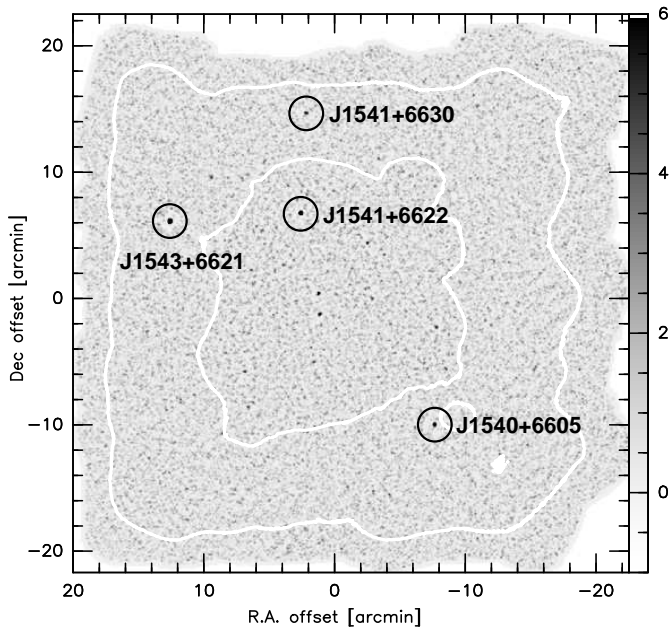}}
\caption{The MAMBO signal to noise 250 GHz map centered on Abell~2125. 
The rms noise level is below 1~mJy per 11 arcsec beam in the
inner contour (ca. 400~arcmin$^2$), and below 3~mJy within the outer
contour (ca. 1200~arcmin$^2$).}
\label{AbellMap}
\end{figure}

Using the Max-Planck Millimeter Bolometer (MAMBO, \cite{Kreysa98, Kreysa02})  1.2~mm (250
GHz) 37- and 117-element arrays at the IRAM 30~m telescope we have imaged four fields: the
Abell~2125 field covers an area of about 1600~arcmin$^2$ (Fig.~\ref{AbellMap}),
the COSMOS field covers $\sim$500~arcmin$^2$, the Lockman Hole covers
$\sim$400~arcmin$^2$ and the NTT deep field covers $\sim$200~arcmin$^2$, adding
up to 2700~arcmin$^2$ imaged to an rms noise level ranging from 0.5 to 3 mJy per
11 arcsec beam. More than 60 significant ($>4\sigma$) sources have been
detected, most of which show a faint ($<100\mu$Jy) 1.4 GHz radio counterpart
(Bertoldi et al., in preparation).

\begin{figure}
\resizebox{\hsize}{!}{\includegraphics{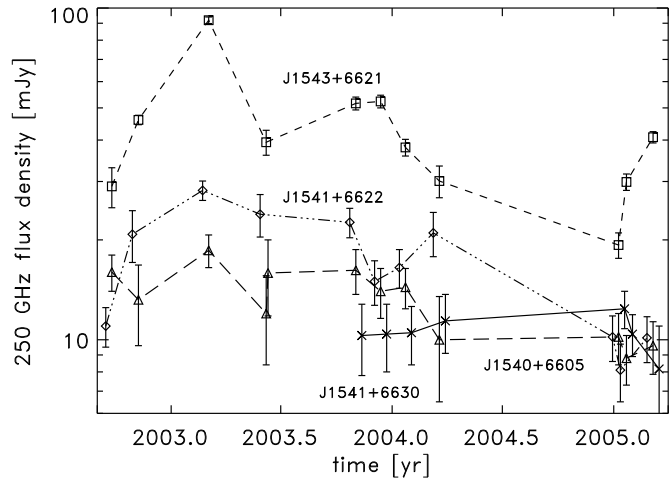}}
\caption{250~GHz flux density measurement of the four brightest sources in the
MAMBO fields, plotted as a function of time. The measurements are
slightly offset horizontally for clarity.}
\label{FluxTime}
\end{figure}

\begin{figure}
\resizebox{\hsize}{!}{\includegraphics{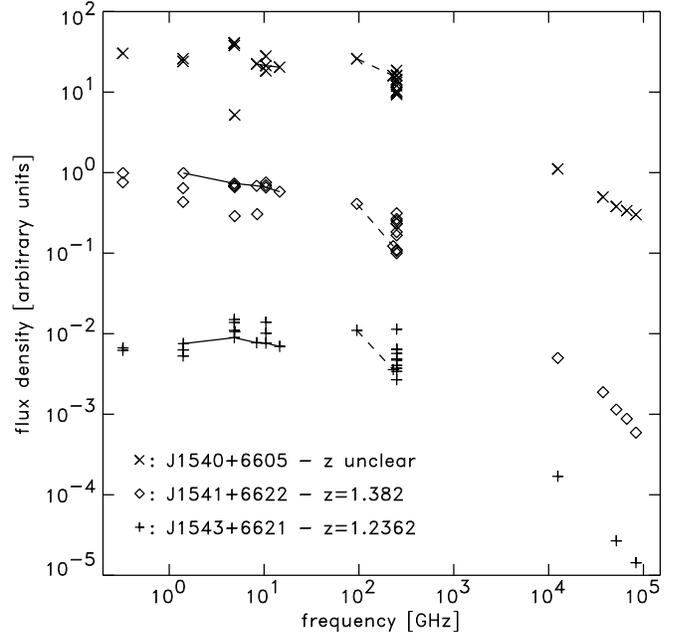}}
\caption{The radio-to-IR SEDs of the three brightest
MAMBO sources, arbitrarily scaled for clarity. The dashed lines connect the simultaneous PdBI measurements (90
and 230~GHz). The solid lines connect near-simultaneous Effelsberg 
measurements (below 20~GHz). The Effelsberg measurements were taken during 
the same day, except for the 1.4~GHz measurements, which were taken two
month earlier. For J1540+6605 measurements below 8~GHz were not taken because 
of confusion with a neighbouring source.}
\label{SEDs}
\end{figure}

We discovered four previously unknown sources with $S_{1.2\mathrm{mm}}\geqslant10$~mJy in
this MAMBO survey (Tables~\ref{Fluxes} and \ref{FluxX}). They are all located in the
Abell~2125 field, which is centered on the diffuse, low density galaxy cluster
Abell~2125 at $z=0.25$ \citep{Owen99}. Their mm flux densities were monitored
with MAMBO pointed observations, showing considerable variability on timescales
of months for the three brighter sources MMJ1540+6605, MMJ1541+6622 and
MMJ1543+6621 (Fig.~\ref{FluxTime}). Those were also observed simultaneously at
1~mm and 3~mm with the Plateau de Bure interferometer (PdBI), showing them to
have spectra declining with frequency (Fig.~\ref{SEDs}).

Deep ($6.5$~$\mu$Jy rms) VLA 20~cm images \citep{Owen05a} show likely
radio counterparts to all four sources, providing positions to about 0.5~arcsec
accuracy. Short observations were also made a 4.86 and 8.46~GHz in August 2000
with the VLA D configuration, which are reported by \cite{Owen05c}, and 324~MHz
observations were made in April 2002 (Owen \& Clarke, private
communication). We performed near-simultaneous cross-scan observations of
MMJ1540+6605, MMJ1541+6622 and MMJ1543+6621 with the Effelsberg 100~m telescope
at 4.85, 8.35, 10.45 and 14.6~GHz on March 24$^{\rm th}$, 2005, and cross-scans
at 1.41~GHz were taken at Effelsberg two month earlier. These near-simultaneous
observations show flat radio spectra. Further cross-scan measurements were taken
at 4.85 and 10.45~GHz nine and eight month earlier, showing radio fluxes that
are variable on timescales of months for all three sources.

We found 1.4~GHz flux measurements for the three brighter sources also in the
NVSS catalog \citep{NVSS}, for two of them at 325 MHz in the WENSS catalog
\citep{WENSS}, and at 5~GHz in the 87GB catalog \citep{87GB}. 


All four sources are covered by our mapping of the Abell~2125 MAMBO field with
the Spitzer Space Telescope, using the Multiband Imaging Photometer for Spitzer
(MIPS) in photometry mode at 24~$\mu$m (rms $\sim 11$~$\mu$Jy) and the Infrared
Array Camera (IRAC) at 3.6~$\mu$m (rms $\sim 0.5$~$\mu$Jy) and 5.8~$\mu$m (rms
$\sim 3$~$\mu$Jy). Three of the sources are also covered by the IRAC 4.5 and
8.0~$\mu$m (rms $\sim 0.6$ and 2.5~$\mu$Jy) parallel observations. The basic
calibrated data have been further reduced using the mopex software package. A
median filter was applied to substract the background from the images. The
source flux densities were measured by fitting a point response function that
was determined from bright sources in the field. MMJ1540+6605, MMJ1541+6622 and
MMJ1543+6621 were detected in all bands, while MMJ1541+6630 was not detected in
any band. 


Spectroscopic redshifts for MMJ1540+6605 and MMJ1541+6630 are not available.
\cite{Miller04} report that MMJ1541+6622 has a redshift of 1.382 based on the
C~III] and Mg~II lines. These broad lines indicate that the object is a quasar.

\begin{figure}
\resizebox{\hsize}{!}{\includegraphics{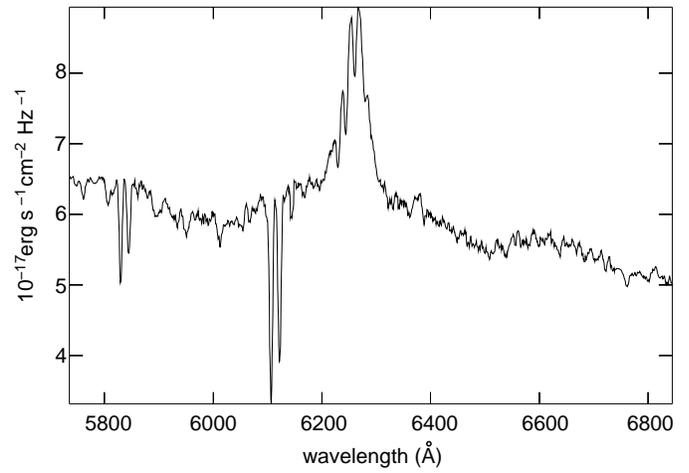}}
\caption{The GMOS/Gemini-N spectrum of MMJ1543+6605, showing the Mg~II line 
at $z=1.2362$.}
\label{spectrum}
\end{figure}

A long slit spectrum of MMJ1543+6605 was obtained using Gemini-N and GMOS on
June 30$^{\rm th}$, 2003 for a total exposure of 30 minutes. The B600 grating
was used which produced a spectrum with about 3.3\AA\ resolution. The spectrum
shows one strong wide emission line (Fig.~\ref{spectrum}) with a number of
associated pairs of absorption lines. This combination and the wavelength
separation of the absorption doublets clearly indicates that the line is Mg~II. 
The redshift we derive is 1.2362. Also the low level emission near 6600~\AA\ is
very consistent with the Fe~II emission line complex at the same redshift.

\begin{figure}
\resizebox{\hsize}{!}{\includegraphics{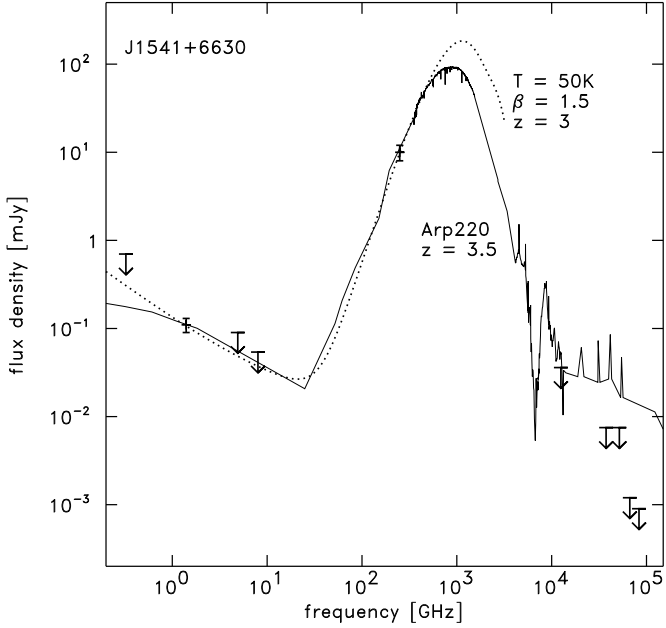}}
\caption{The radio-to-IR SED of MMJ1541+6630. The
dotted line shows a greybody with $T_\mathrm{dust}=50$~K and $\beta=1.5$ and a
synchrotron spectrum with $\alpha=0.7$ at $z=3$, whose radio-to mm flux ratio
is set according to the correlation found in local galaxies \citep{Carilli99}.
The solid line is the SED of Arp~220 shifted to $z= 3.5$ and scaled to match the radio
and mm measurements.}
\label{OneSED}
\end{figure}

We estimate a photometric redshift of 3$_{-1.2}^{+1.9}$ for MMJ1541+6630
(Fig.~\ref{OneSED}), using the correlation between the radio and FIR luminosities
found in the local star-forming galaxies \citep{Condon92, Carilli99, Carilli00}. 

\begin{figure}
\resizebox{\hsize}{!}{\includegraphics{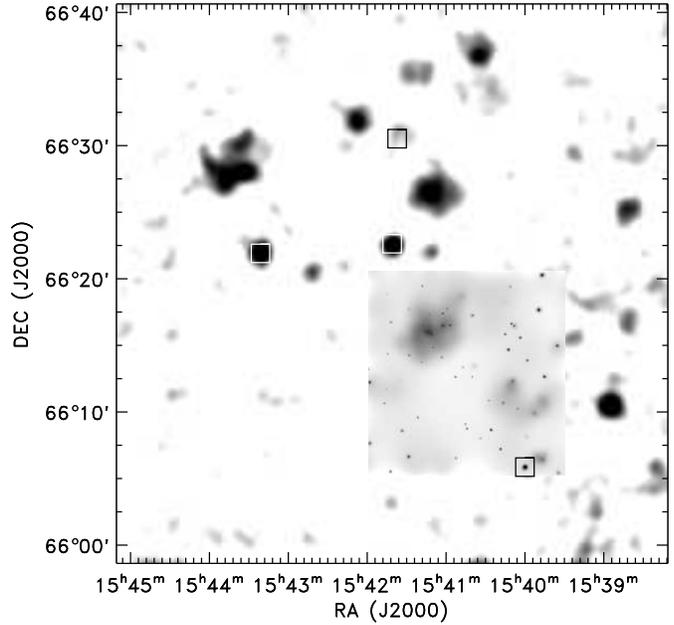}}
\caption{Smoothed ROSAT PSPC map of the Abell 2125 field with inlayed
smoothed Chandra ACIS-I map. The logarithmic grey scale ranges from 
$1\cdot10^{-3}$~ct/s to $6\cdot10^{-3}$~ct/s in the ROSAT image and from 
$1\cdot10^{-4}$~ct/s to $8\cdot10^{-3}$~ct/s in the Chandra inlay. The
positions of the four bright MAMBO sources are indicated by squares.}
\label{Rosat} 
\end{figure}

We found X-ray counterparts for MMJ1540+6605, MMJ1541+6622 and MMJ1543+6621 in
the ROSAT 1WGA catalog \citep{WGAcat}. MMJ1541+6630 is detected in a 20~ksec
ROSAT PSPC observation. MMJ1540+6605 is detected in a 83~ksec archival Chandra
observation (Fig.~\ref{Rosat}; \cite{Wang04a}). The 0.5-10~keV flux measured by
Chandra is $(1.1\pm0.1)\cdot10^{-13}$~erg~cm$^{-2}$~s$^{-1}$. 

The offsets of all identifications to the VLA positions are less than the
$1\sigma$ positional error quoted in the catalogs, except for the NVSS
counterpart of MMJ1543+6621 and the ROSAT counterpart of MMJ1541+6630, for
which the offsets are within the $2\sigma$ positional error.

\begin{table*}
\begin{center}  
\caption{\label{Fluxes}The radio to mm source properties of the four
brightest MAMBO sources}
\begin{tabular}{lllllllllll}
\hline \hline
{\bf Source (J2000)}     & {$S_{0.3}$} & {$S_{1.4}$} & {$S_{4.9}$} & {$S_8$} & {$S_{10}$}&{$S_{15}$}& {$S_{95}$}  & {$S_{250 \mathrm{GHz}}$} & $z$ & $\alpha^{230}_{95}$\\
                         & [mJy]       & [mJy]	     & [mJy]	   & [mJy]   & [mJy]     & [mJy]    &  [mJy]      & [mJy]                    &     &                    \\
\hline
{\bf 154000.01+660551.6} & $30.3\pm0.2$&24--26       & 5--41       &$22\pm1$ & 21--28    & $20\pm2$ & 26$\pm$1    & 12--19                   & unclear & $0.5\pm0.3$\\
{\bf 154137.19+663031.3} & $<0.7$      &0.10$\pm$0.02& $<0.09$     &$<0.054$ &           &          &             & 10$\pm$2                 &3$_{-1.2}^{+1.9}$& \\
{\bf 154141.01+662237.9} & 68--89      & 39--89      & 26--66      & 28--62  & 59--68    & $52\pm2$ & 37$\pm$2    & 11--28                   &1.382   &$1.4\pm0.3$\\
{\bf 154321.31+662154.5} & $50.3\pm0.4$& 43--61      & 72--121     & $63\pm1$& 62--112   & $56\pm1$ & 89$\pm$5    & 29--92                   &1.2362   &$1.3\pm0.2$\\
\hline
\end{tabular}
\\[2pt]
\end{center}
Origin of the measurements: 0.3~GHz: VLA \& Westerbork; 1.4 \& 8~GHz: VLA, Effelsberg;
4.9 GHz: Effelsberg, VLA \& Green Bank 300ft; 10 \& 15~GHz: Effelsberg; 95~GHz: PdBI; 250~GHz: PdBI \& MAMBO\\
The spectral index $\alpha$ is defined as $S_\nu\propto\nu^{-\alpha}$ in
this paper.
\end{table*}

\begin{table*}
\begin{center}  
\caption{\label{FluxX}The mid-infrared to X-ray properties of the four brightest MAMBO sources}
\begin{tabular}{llllllllll}
\hline \hline
{\bf Source (J2000)}     & $S_{24}$ & $S_{8}$  &$S_{5.8}$&$S_{4.5}$&$S_{3.6\mu{\rm m}}$& R     & 0.1--2.4 keV            & Hardness  & {$L_{0.1-2.4{\mathrm{keV}}}$}\\
                         &[$\mu$Jy] &[$\mu$Jy] &[$\mu$Jy]&[$\mu$Jy]& [$\mu$Jy]         & [mag] & [erg cm$^{-2}$ s$^{-1}$]&           & [erg s$^{-1}$]               \\
\hline
{\bf 154000.01+660551.6} &$1113\pm7$&$498\pm10$&$380\pm8$&$338\pm4$& $301\pm3$         & 20.9  & $(4\pm.6)\cdot10^{-14}$ &$1.0\pm.4$ & $(4.2\pm.6)\cdot10^{43}$	    \\
{\bf 154137.19+663031.3} & $<36$    & $<7.5$   & $<7.5$  & $<1.2$  & $<0.9$            &$>27.2$& $(3\pm.5)\cdot10^{-14}$ &$0.4\pm.3$ & $(2.4\pm.4)\cdot10^{45}$	    \\
{\bf 154141.01+662237.9} &$450\pm5$ &$169\pm8$ &$103\pm5$& $79\pm2$& $53\pm2$          & 20.7  & $(8\pm.8)\cdot10^{-14}$ &$0.6\pm.3$ & $(9.4\pm.9)\cdot10^{44}$     \\
{\bf 154321.31+662154.5} &$1371\pm18$&          &$217\pm6$&         & $116\pm2$         & 19.3  & $(1\pm.1)\cdot10^{-13}$ &$0.4\pm.3$ & $(8.9\pm.7)\cdot10^{44}$     \\
\hline
\end{tabular}
\\[2pt]
\end{center}
Origin of the measurements: 24~$\mu$m - 3.6~$\mu$m: Spitzer MIPS \& IRAC; R: KPNO; 0.1--2.4~keV: ROSAT PSPC\\
Hardness: ROSAT HR1 is defined as ($M+J-C$)/($M+J+C$) where $M, J, C$
are the countrates in the 0.5--1.3, 1.5--2.1 and 0.1--0.3~keV bands.\\
For the luminosities we assumed $\Omega_M=0.27$, $\Omega_\Lambda=0.73$ and 
$H_0=71$~km~s$^{-1}$~Mpc$^{-1}$. For MMJ1540+6605 we assumed a redshift
of 0.5, which is suggested by optical photometry.
\end{table*}


\section{Discussion}


To quantify the significance of our source identifications, we calculated the
corrected Poissonian probability, $P$, of a chance association
\citep{Browne78, Downes86}. For the identifications from the NVSS, WENSS and 87GB
catalogs, $P<0.1$\%, so that the found counterparts correspond to the MAMBO
sources with $>99.9$\% confidence. For the ROSAT identifications the confidence
is 93\%, 83\%, 98\% and 99\% for MMJ1540+6605, MMJ1541+6630, MMJ1541+6622, and
MMJ1543+6621, respectively. 

As the absorption of X-rays by hydrogen decreases with increasing photon energy,
the hardness ratios can be used for rough estimates of the absorbing column
densities. Three of the sources show absorption by
$N_\mathrm{H}\sim1-4\cdot10^{20}$~cm$^{-2}$, comparable to the galactic value of
$2.4\cdot10^{20}$~cm$^{-2}$ toward the Abell~2125 field estimated from the
Leiden/Dwingeloo survey \citep{Hartmann97}. For MMJ1540+6605 we estimate 
$N_\mathrm{H}\approx(2\pm0.4)\cdot10^{21}$~cm$^{-2}$, which hints at intrinsic
absorption. 

\subsection{MMJ1541+6630: A thermal source}

MMJ1541+6630 has a faint 1.4~GHz radio counterpart with a flux density typical
of the MAMBO or SCUBA galaxies. The 1.4~GHz flux density measurement and the
upper limits at 4.9 and 8~GHz indicate a falling radio spectrum, consistent with
a typical synchrotron spectrum with $\alpha\sim0.7$
($S_\nu\propto\nu^{-\alpha}$). The ratio between the 1.4 and 250~GHz flux
densities is consistent with a typical starburst galaxy SED at $z\sim3$. No
optical counterpart could be identified down to an R magnitude of 27.2, which is
not unusual for even the brightest mm galaxies, e.g.\ MMJ120456-0741.5
\citep{Dannerbauer02, Dannerbauer04}. The standard interpretation of these data
is that this object is a luminous high redshift starburst. 

In starforming galaxies in which no AGN
contributes significantly to the X-ray emission the radio and the FIR
luminosity are also found to be well correlated with the X-ray luminosity
\citep{Ranalli04}. The X-ray source close to the radio position of
MMJ1541+6630 is detected at a low signal to noise ratio, making the
position  determination uncertain. The confidence of an association of the
X-ray source and MMJ1541+6630 is 83\%. In the following we assume that the
ROSAT source is associated with MMJ1541+6630. The inferred X-ray
luminosity is $\sim80$ times higher than that
expected from the correlations between X-ray, radio and mm
luminosities for starforming galaxies. This
X-ray excess reveals the presence of a radio-quiet AGN, which is
also not unusual for (sub)mm galaxies: \cite{Alexander04} suggest that
$\sim40$\% of the (sub)mm galaxies host AGN, which however contribute less
then 20\% to the IR bolometric luminosity. Following the classification
scheme suggested by \cite{Hasinger03}, the implied X-ray luminosity
classifies MMJ1541+6630 as a quasar. 

Taking the SED of the local ULIRG Arp~220, redshifted to $z=3.5$, and fitting it to
the mm and radio measurements of MMJ1541+6630 would yield flux densities in the
Spitzer IRAC bands that are 3.5 to 20 times higher than the measured IRAC upper
limits. A typical ULIRG SED with emission from hotter dust ($T>50$~K) than that in
Arp~220, located at higher redshifts than 3 would still match the radio and mm
measurements \citep{Blain99a} but would yield lower flux densities in the IRAC
bands. It could be that MMJ1541+6630 is therefore at $z>3$.

Scaling the average SED for a radio-quiet quasar from \cite{Elvis94} to match
the observed X-ray flux, we would expect MMJ1541+6630 to have an R magnitude of
20.5, much brighter than what is observed. The expected flux densities in the
Spitzer IRAC and MIPS 24~$\mu$m bands would also exceed the measurements by an
order of magnitude. Dust obscuration by $\geqslant6.7$ magnitudes at
$\lambda=0.16$~$\mu$m (rest frame) could explain the optical faintness. Assuming
galactic dust properties this would translate to an $A_V\geqslant2.6$. To match
the flux densities observed in the Spitzer bands,  with galactic dust properties
$A_V$ needs to be of order $50-100$, which is not atypical for starburst
galaxies; e.g.\ in Arp~200 $A_V\sim50$. Assuming a galactic gas to dust ratio of
150, $A_V\geqslant50$ would imply a hydrogen column density $N_{\rm
H}\geqslant9\cdot10^{22}$~cm$^{-2}$, which is about two orders of magnitude
larger than the value inferred from the X-ray hardness. Due to the low signal to
noise ratio of the ROSAT measurement, the X-ray inferred column density is
uncertain by at least a factor of five. Another uncertainty arises from the high
redshift, whereby the soft part of the X-ray spectrum, which is most affected by
absorption, is shifted out of the ROSAT bandpass, so that much higher
column densities would still be consistent with the observed hardness ratio (see
e.g.\ Alexander et al. 2004). The difference in
inferred column densities might also hint on different line of sights
toward the starburst and the quasar.

Studies of quasars at mm wavelengths show almost no correlation between the
optical emission of quasars and their mm emission \cite[ and references
therein]{Omont03}. 
Radio quiet quasars studied at cm and (sub)mm wavelengths show the same
ratio of cm and (sub)mm inferred luminosities as known from starforming
galaxies (Petric et al.\ in preparation; Beelen et al.\ in preparation).
Besides this indirect evidence for a significant, if not
dominant dust heating by a starburst in quasars, for the high redshift
quasars 
PSS~J2322+1944 
\citep{Carilli03},
SDSS~J114816.64+525150.3
\citep{Bertoldi03b} and the Cloverleaf \citep[H1413+117,][]{Weiss03} a
starburst could directly be identified as the dominant dust heating source.
The Spitzer measurements of MMJ1541+6630 rule out a significant
contribution to the dust heating by an AGN, except if this source were a
heavily obscured $z\sim 5$ quasar with moderate star forming activity.

With these uncertainties in
mind, we find the properties of MMJ1541+6630 consistent with typical
X-ray luminous (sub-)mm galaxies such as those studied by Alexander et al.
(2004). In the following we shall assume that MMJ1541+6630 is a quasar with
its FIR emission dominated by star formation.

For a dust temperature of 50~K and an emissivity index $\beta=1.5$, typical for
local starburst galaxies \citep{Dunne01}, at $z=3$ MMJ1541+6630 would have an
infrared luminosity of $(4\pm1)\cdot10^{13}$~L$_\odot$, making it one of the
most luminous starbursts known. It would be more luminous in the (sub)mm
and X-rays than the type-1 quasar SMMJ04135 that was recently discovered in
SCUBA blank fields \citep{Knudsen03}, or the type-2 quasar SMMJ02399
\citep{Ivison98, Genzel03}. Note that the
relation between 250~GHz flux density and infrared luminosity is nearly
independent of redshift from $z\approx0.5$ to 10. 

With a Salpeter initial mass
function and a starburst age of $\sim50$~Myr \citep{Omont01} the inferred star
formation rate would be 4500~M$_\odot$~yr$^{-1}$. Such a high star formation
rate would imply a very massive object. Adopting a dust emissivity
$\kappa_\mathrm{125\mu m}=18.75$~cm$^2$~g$^{-1}$ \citep{Hildebrand83} the
inferred dust mass of MMJ1541+6630 is $\sim 1\cdot10^{9}$~M$_\odot$. 

The likely detection of a massive starburst associated
with a quasar adds to the growing evidence that the growth of central
supermassive black holes and the formation of the first stellar populations
are closely linked \citep[e.g.][ and references therein]{Haehnelt00}. Today
we see the fossil records of this connection in the correlation between
black hole and stellar bulge masses in local galaxies \citep{Magorrian98,
Ferrarese00, Gebhardt00a}. Following \cite{Omont01}, for MMJ1541+6630 we
estimated the ratio between star formation rate, $\dot{M}_\mathrm{SF}$, and
black hole growth rate, $\dot{M}_\mathrm{BH}$, as well as the bulge to
black hole mass ratio, $M_\mathrm{bul}/M_\mathrm{BH}$. The former is
estimated by relating the FIR luminosity to the bolometric luminosity
arising from the quasar, which we calculate from the X-ray luminosity using
the average radio-quiet quasar SED from \cite{Elvis94}. For the latter
estimate we assume the dust properties mentioned above. 

We find $\dot{M}_\mathrm{SF} / \dot{M}_\mathrm{BH}\approx$~3400. This is
$\sim10$ times higher than expected if black holes and stars formed
simultaneously with the same ratio of bulge mass to black hole mass as observed
in local galaxies. For SMMJ02399 we find $\dot{M}_\mathrm{SF} /
\dot{M}_\mathrm{BH}\approx$~2200, still a factor of $\sim6$ higher. For
$M_\mathrm{bul}/M_\mathrm{BH}$ we find $\sim$~1200 and $\sim850$ for
MMJ1541+6630 and SMMJ02399 respectively, a factor of $\sim2-3$ above the value
found in local galaxies. Given the uncertainties in our
$M_\mathrm{bul}/M_\mathrm{BH}$ estimate and the scatter in
$M_\mathrm{bul}/M_\mathrm{BH}$ for local galaxies, the estimated ratios are
similar to those in local galaxies. Compared with the optically selected
quasars of \cite{Omont01}, the bolometric luminosities of the (sub)mm selected
quasars MMJ1541+6630 and SMMJ02399 appear to be dominated by the starburst
activity.

\subsection{Flat spectrum sources}

The flat and time-variable radio to mm SEDs of the three brightest sources, their
relatively bright optical counterparts and their high X-ray luminosities indicate
that these sources are flat spectrum, radio loud quasars. The broad emission
lines in the optical spectrum of MMJ1541+6622 also indicate that this source is a
quasar. The Spitzer SEDs of MMJ1541+6622 and MMJ1543+6621 can well be fit with
the average SEDs of radio loud quasars \citep{Elvis94} shifted to the
spectroscopically measured redshifts. The 95 to 230~GHz spectral index measured
with the PdBI also suggests a non-thermal spectrum, which is however steeper by
$\sim0.5$--1 than the radio spectrum between 1 and 95~GHz. The fact that the
radio spectrum is flat up to 95~GHz indicates that the steepening occurs very
close to this frequency. A similar mm-break is also seen e.g.\ in 3 quasars
selected from the 3C catalog \citep{VanBemmel01}.


Flat spectrum radio sources (FSS) are well studied at radio
wavelengths, but little is known about their SEDs at frequencies above
90 GHz and below the MIR.  To see how the mm bright FSS relate to the
radio selected FSS, we used the 90~GHz observations of \cite{Holdaway94}, as
interpreted by the model of \cite{Holdaway05}, to extrapolate the source
counts up to 250~GHz.  This model uses 90~GHz observations of flat spectrum
quasars selected at cm wavelengths to sample the spectral index distribution
of the sources' compact cores from 8.4~GHz to 90~GHz.  To push the model to
frequencies above 90~GHz, Holdaway and Owen determined plausible distributions
of turnover frequencies in these FSS, assuming the spectral index turns over to
0.8 above these frequencies.

The sampling of the spectral index distribution of the FSS population
up to 90~GHz builds on the work of \cite{Patnaik92}, who selected
sources with $S_{5 GHz}>200$~mJy and 1.4 to 5~GHz spectral index
$\alpha < 0.5$.  The 1.4~GHz data \citep{Condon85, Condon86}
and the 5~GHz data \citep{Condon89} that defined their survey
were taken with the NRAO 91~m telescope. \cite{Patnaik92}
observed their FSS sample with the VLA A array at 8.4~GHz.
\cite{Holdaway94} defined a subsample of Patnaik's sample,
selecting 367 sources with 8.4~GHz core fluxes ranging from 200~mJy to
35~Jy. \cite{Holdaway94} observed these sources with the NRAO
12~m telescope.  As the 8.4~GHz observations with 0.2 arcsecond
resolution provided excellent core fluxes, and from 8.4~GHz to 90~GHz,
the steep spectrum emission is decreased by a factor between 10 and
100, a good spectral index distribution for the FSS cores between 
8.4~GHz and 90~GHz was obtained.  No evidence was found for a
spectral index distribution depending on core flux.

\cite{Condon84} presents source counts of FSS at 5~GHz, complete
down to 10 mJy, from a variety of single dish observations.
These counts, which included contributions from both a flat spectrum
core and steep spectrum extended emission, were reduced by a mean
core fraction of 0.8 and then scaled up to 90~GHz with the
8.4~GHz to 90~GHz spectral index distribution.  To scale the
FSS counts to other frequencies, we need a model for how the
spectral index varies at higher frequency.  Such a model is hinted
at by the distribution of spectral index between 8.4~GHz and 90~GHz,
which can be decomposed into two Gaussian components, indicating
that 81\% of the observed sources have ``turned over'' to a steep
spectral index by 90~GHz, and 19\% of the observed sources remain flat
at 90~GHz.  By combining the theoretical knowledge that flat
spectrum quasars have a spectral index of about 0.0, and that steep
spectrum, optically thin synchrotron emission has a spectral index of
about 0.8, and by making reasonable assumptions on the spread
about those spectral index values, \cite{Holdaway05} were
able to solve for a distribution of turn over frequencies.
This distribution of turn over frequencies can then be combined
with the well-studied 5~GHz FSS counts and the core fraction correction 
to estimate the FSS counts at any frequency above 8.4~GHz.

One thing remains undetermined in this model: what happens to those
19\% of still-flat spectrum sources above 90~GHz?  An upper limit to
source counts at 250~GHz can be obtained by assuming that these sources
remain flat spectrum through 250~GHz.  A lower limit can be obtained
by assuming they all turn over right at 90~GHz.  A ``best guess'' can
be obtained by continuing the best fit power law distribution of
turn over frequencies up to 181~GHz, at which point all of the
remaining flat spectrum sources have turned over.

\begin{figure}
\resizebox{\hsize}{!}{\includegraphics{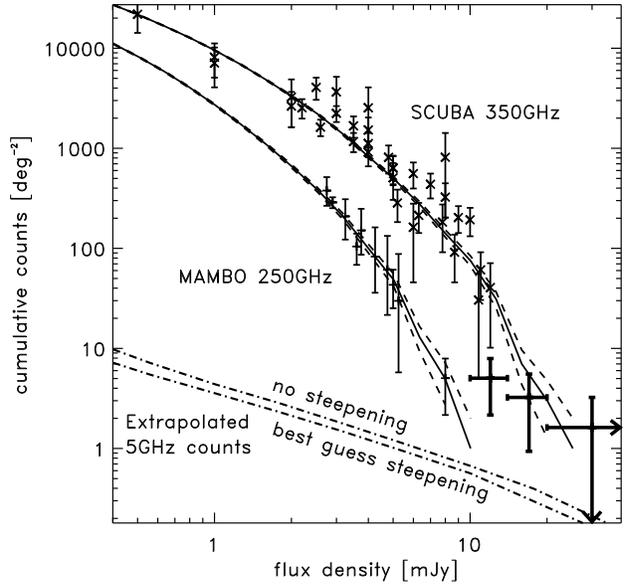}} 
\caption{The cumulative source counts based on the MAMBO and SCUBA surveys.
The rightmost three data points correspond to the FSS and were
not included in the counts below 10 mJy. The lines show the model fit
(solid) of the thermal population's counts and its 1$\sigma$-error
(dashed). The dotted lines represent the extrapolation of the 5~GHz FSS
counts using the model of \cite{Holdaway05}.} 
\label{Counts} 
\end{figure}

The extrapolated number counts are shown in Fig.~\ref{Counts}, with no
spectral steepening up to 250~GHz and with the ``best guess'' model for
spectral steepening. The FSS source counts from the MAMBO fields are a
factor of $9^{+33}_{-3}$ or $7^{+30}_{-3}$ higher than predicted from the
extrapolation with and without steepening, respectively. Lensing
amplification is unlikely to be a cause for the observed source overdensity
since the four sources are far from the inner $2\times2$~arcmin region of
the galaxy cluster where mild lensing amplification could be expected. It
could be that the number of FSS in the MAMBO field is high by chance.
With Monte Carlo methods we simulated a field of
1000 square degrees with source counts that are consistent with the
model and sources at random positions. We took 1600 independent
small fields of the MAMBO survey size from the large simulation and
analysed the distribution of source counts for the small fields. We found that less
than 0.4\% of these artificial fields contain three or more sources brighter than 10~mJy. This discrepancy
might also be due to a deviation from the found spectral index distribution
for sources below the 90~GHz detection limit of 75~mJy. Although the
spectral index distribution was found not to depend on radio flux density,
this is based on the brighter sources ($S_\mathrm{5GHz}>200$~mJy) that were
also detected at 90~GHz. It is possible that some fainter sources have
flatter or more inverted spectra and thus contribute noticeably to the mm
number counts.


The four sources are among the first quasars to be discovered in mm or submm
blank field surveys. Although based on small number statistics, they allow a
first study of the contribution of non-thermal sources in (sub)mm observations.
Figure~\ref{Counts} shows the cumulative source counts for the MAMBO and SCUBA
deep field surveys. We have fitted the counts with a model of a population of
dust enshrouded starburst galaxies. The model is based on the Saunders
60~$\mu$m luminosity function of local galaxies \citep{Saunders90} which
consists of a flat power law for low luminosities and a steep power law for
luminosities above $\sim10^9$~L$_\odot$. An additional exponential cutoff at
$\sim10^{12}$~L$_\odot$, and a luminosity and density evolution with redshift
were applied to match the (sub)mm number counts and their redshift
distribution. The model was found to be consistent with the SED of the
extragalactic FIR background. Including MMJ1541+6630 in the number counts, we
find that they are in good agreement with the extrapolation of the fainter
source counts. The three brightest sources however indicate a flattening in the
source counts at 250~GHz flux densities $>10$~mJy. The mm background appears to
change from a declining population of high redshift starbursts to a shallower
distribution of less distant, radio loud quasars.

\begin{acknowledgements} 

We thank J\"urgen Kerp from RAIUB and Thomas Boller from MPE Garching for their
help with the ROSAT data and Matthias Kadler from the MPIfR for his help with the
Chandra data. We thank Endrik Kr\"ugel from MPIfR for his useful comments on
optical dust properties,  Emmanouil Angelakis and Alex Kraus from the MPIfR for
their help on Effelsberg observations, and  the IRAM staff for their great
support of this project.
CC acknowledges support from the Max-Planck-Forschungspreis, granted by the
Alexander von Humboldt Foundation and the Max-Planck Society. HV is member of the
International Max Planck Research School (IMPRS) for Radio and Infrared Astronomy.

This research has made use of the NASA/IPAC Extragalactic Database
(NED) which is operated by the Jet Propulsion Laboratory, California
Institute of Technology, under contract with the NASA, and of data obtained
through the High Energy Astrophysics Science Archive Research Center
Online Service, provided by the NASA/Goddard Space Flight Center.
\end{acknowledgements}
\bibliographystyle{aa}
\bibliography{./3712voss}
\end{document}